\documentclass[twocolumn,showkeys,showpacs,preprintnumbers,prd,superscriptaddress,nofootinbib,aps,10pt]{revtex4-1}
\usepackage{graphicx,epsf,bm,amsmath,amsfonts,amssymb,epstopdf,natbib,hyperref,color,verbatim,multirow,bm,appendix}
\hypersetup{colorlinks=true,urlcolor=blue,citecolor=blue,linkcolor=blue,menucolor=blue,anchorcolor=blue,filecolor=blue}
\date{\today}

\begin{document}

\title{Probing the small-scale primordial power spectrum\\via relic neutrinos and acoustic reheating}

\author{Giovanni Piccoli}
\email{giovanni.piccoli@uzh.ch}
\affiliation{Department of Astrophysics, University of Zurich, Winterthurerstrasse 190, CH-8057, Zurich, Switzerland \looseness=-1}

\author{Sunny Vagnozzi}
\email{sunny.vagnozzi@unitn.it}
\affiliation{Department of Physics, University of Trento, Via Sommarive 14, 38123 Povo (TN), Italy}
\affiliation{Trento Institute for Fundamental Physics and Applications (TIFPA)-INFN, Via Sommarive 14, 38123 Povo (TN), Italy}

\author{Joseph Silk}
\email{silk@iap.fr}
\affiliation{Institute d'Astrophysique de Paris (IAP), 98bis Boulevard Arago, F-75014 Paris, France \looseness=-1}
\affiliation{Department of Physics and Astronomy, The Johns Hopkins University, 3400 N.\ Charles Street, Baltimore, MD 21218, USA \looseness=-1}
\affiliation{BIPAC, Department of Physics, University of Oxford, Keble Road, Oxford OX1 3RH, United Kingdom \looseness=-1}

\begin{abstract}
\noindent We show that the dissipation of small-scale perturbations through diffusion damping after neutrino decoupling lowers the present-day neutrino temperature compared to the expected value of $1.96\,{\text{K}}$. This reduces the relic neutrino abundance by an amount controlled by the integral of the primordial curvature power spectrum $\Delta_{\cal R}^2(k)$. We find that a relic neutrino detection by PTOLEMY can set limits $\Delta_{\cal R}^2(k) \lesssim {\cal O}(0.1)$ on scales $k \lesssim 3 \times 10^5\,{\text{Mpc}^{-1}}$, complementary to limits from Big Bang Nucleosynthesis, spectral distortions, pulsar timing arrays, and future dark ages 21-cm observations.
\end{abstract}

\maketitle

\textbf{\textit{Introduction.}} --- On scales $k \gtrsim {\cal O}({\text{Mpc}}^{-1})$, well beyond those probed by the Cosmic Microwave Background (CMB) and large-scale structure (LSS), primordial perturbations are poorly constrained~\cite{Ballardini:2016hpi,Planck:2018jri,SimonsObservatory:2018koc,SimonsObservatory:2019qwx,Ballardini:2022wzu,Euclid:2023shr,Antony:2024vrx,DESI:2024hhd,Raffaelli:2025kew,Chudaykin:2025vdh}. However, access to these modes would allow us to probe various physical processes of interest, such as additional fields during inflation, features associated with the formation of primordial black holes (PBHs), or effects beyond General Relativity~\cite{Achucarro:2010da,Pajer:2013fsa,Odintsov:2017qpp,Nojiri:2017ncd,Green:2020jor,Villanueva-Domingo:2021spv,Bird:2022wvk,Carr:2023tpt,Odintsov:2023weg,Choudhury:2024aji}. Extending constraints on primordial perturbations beyond the narrow CMB--LSS window would also help in distinguishing between different inflationary models, and reconstructing the inflaton potential. In principle these small-scale modes survive in the LSS, but non-linear evolution and baryonic effects make it challenging to recover their primordial properties. In the CMB, the same modes are instead damped before recombination by photon diffusion, a process first identified by one of us and referred to as \textit{Silk damping}~\cite{Silk:1967kq}.

Damping erases small-scale modes while injecting energy into the plasma and altering the thermal history of the Universe. Specifically, modes with $k \gtrsim k_D(z)$ dissipate and release energy sourced by primordial perturbations into the plasma, where $k_D(z)$ denotes the diffusion wavenumber~\cite{Sunyaev:1970plh,Daly:1991uob,Barrow:1991hwg,Hu:1994bz,Chluba:2012gq}. At redshifts $z \gtrsim 2 \times 10^6$, this increases the entropy of the photon bath~\cite{Chluba:2011hw,Khatri:2011aj}. The entropy produced during this \textit{acoustic reheating} process changes the thermal history in a predictable way, turning loss of small-scale power into a potential probe of high-$k$ primordial fluctuations~\cite{Jeong:2014gna,Nakama:2014vla,Naruko:2015pva,Inomata:2016uip,Ota:2017jte}. The energy released during acoustic reheating is in fact proportional to the integral of the dimensionless curvature power spectrum $\Delta_{\cal R}^2(k)$. The standard cosmological thermal history therefore implicitly relies on the assumption that $\int d\ln k\,\Delta_{\cal R}^2(k) \ll 1$, approximation which can fail in the presence of enhanced small-scale power, as acoustic reheating causes photons to cool more slowly than the usual $T_{\gamma} \propto 1/a$ scaling. Given that the present-day CMB temperature $T_{\gamma,0} \approx 2.75\,{\text{K}}$ is fixed to extremely high precision~\cite{Mather:1993ij,Fixsen:1996nj,Fixsen:2009ug}, acoustic reheating effectively results in the photon bath being progressively colder at earlier times compared to the usual expectation. As for neutrinos, after decoupling from the plasma at $T \sim 2\,{\text{MeV}}$, they do not share the entropy produced by acoustic reheating, and thus maintain the standard $T_\nu \propto 1/a$ evolution from then on. This lowers the present-day neutrino temperature compared to the canonical value $T_{\nu,0} \approx 1.96\,{\text{K}}$.

Reducing $T_{\nu,0}$ lowers the relic neutrino number density $n_{\nu,0}$. This is a key quantity for experiments aiming to detect the cosmic neutrino background (CNB), such as PTOLEMY~\cite{PTOLEMY:2019hkd}, which can therefore set integral limits on the small-scale primordial power spectrum. This \textit{Letter} develops this connection concretely, studying the impact of acoustic reheating on $n_{\nu,0}$ for a range of well-motivated enhanced small-scale power spectra, and assessing PTOLEMY's potential to probe small-scale fluctuations beyond the CMB--LSS window. We show that relic neutrinos can set ${\cal O}(0.1)$ level limits on small-scale primordial fluctuations on scales complementary to those probed by Big Bang Nucleosynthesis (BBN), spectral distortions (SDs), gravitational waves (GWs), and future dark ages 21-cm observations.

\textbf{\textit{Acoustic reheating and relic neutrinos.}} --- Diffusion damping continuously erases small-scale perturbations in the photon-baryon plasma, converting the energy stored in acoustic modes into the isotropic radiation background (a process referred to as Silk damping in the late-time, photon diffusion regime, but more generally arising from viscosity and thermal conductivity of the primordial plasma). The consequences of this process are easily understood by describing the inhomogeneous photon field as a superposition of local blackbodies (BBs) with direction-dependent temperature $T_{\gamma}(t,\bm{x},\hat{\bm{p}})=T_{\gamma}(t)[1+\Theta(t,\bm{x},\hat{\bm{p}})]$. The spatial average of such a superposition is not a perfect BB: at second order in $\Theta$ it features a relative excess energy density of $\Delta \rho_{\gamma}/\rho_{\gamma} \simeq 2\langle\Theta^2\rangle$ compared to a BB at temperature $T_\gamma(t)$ or, equivalently, a relative deficit in photon number of $\Delta N_{\gamma}/N_{\gamma} \simeq -3\langle\Theta^2\rangle/2$ compared to a BB with the same total energy density~\cite{Chluba:2012gq}. Physically, as small-scale modes enter the horizon they dissipate their energy stored in sound waves through particle diffusion. For $z\gtrsim 2\times 10^6$, i.e.\ prior to neutrino decoupling, efficient thermalization replenishes the missing photons, increasing the comoving photon number $n_{\gamma}$ and the entropy of the photon bath. This process is called \textit{acoustic reheating}.

\begin{figure}
\centering
\includegraphics[width=0.82\linewidth]{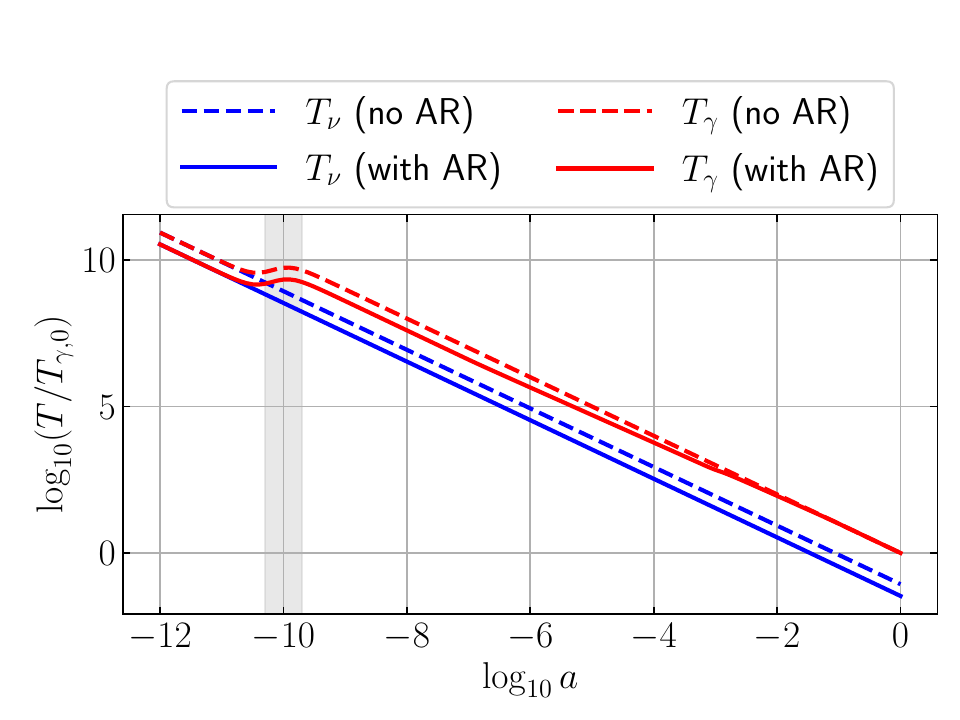}
\caption{Schematic illustration of the impact of acoustic reheating (AR). The photon (red) and neutrino (blue) temperatures are shown as function of the scale factor. Fixing the present-day photon temperature to $T_{\gamma,0}$ (to which both red curves converge), acoustic reheating reduces the present-day neutrino temperature (see solid blue curve). The gray vertical band indicates neutrino decoupling.}
\label{fig:temperature_evolution}
\end{figure}

To understand this, consider the evolution of $n_{\gamma}$~\cite{Chluba:2012gq}:
\begin{equation}
\frac{d}{dt}\ln \left ( n_{\gamma} a^3 \right ) = -\frac{3}{2}\frac{d}{dt}\langle\Theta^2\rangle\,,
\label{eq:ngamma}
\end{equation}
so that dissipation ($d\langle\Theta^2\rangle/dt<0$) increases $n_{\gamma}a^3$, whereas the total energy density is unchanged and merely redistributed. For modes well inside the horizon during the radiation era, the mean-squared temperature anisotropy $\langle\Theta^2\rangle$ is given by~\cite{Chluba:2012gq,Chluba:2013dna}:
\begin{equation}
\langle\Theta^2(a)\rangle \simeq C_\nu^2 \int_0^\infty\frac{dk}{k}\,\Delta_{\cal R}^2(k)\exp \left [-\frac{k^2}{k_D^2(a)} \right ] \,,
\label{eq:theta}
\end{equation}
where $C_{\nu}=(1+4f_{\nu}/15)^{-1}$ accounts for neutrino anisotropic stress, with $f_{\nu}=\rho_{\nu}/(\rho_{\gamma}+\rho_{\nu})$ being the fraction of the total radiation energy density carried by neutrinos, and $k_D(a)$ is the diffusion wavenumber, above which ($k \gtrsim k_D$) modes are damped. As the Universe expands, $k_D$ decreases (the damping scale $\propto k_D^{-1}$ increases), so progressively larger scales are being damped. Using the subscript $_0$ to denote present-day values, let us define the quantity $\delta\vartheta(a)\equiv[\langle\Theta^2(a)\rangle-\langle\Theta^2(a_0)\rangle]/2$, which is zero today ($a=a_0$), and positive in the past. Then, Eq.~(\ref{eq:ngamma}) can be integrated to give $n_\gamma(a)= n_{\gamma,0}a^{-3}\exp[-3\delta\vartheta(a)]$, from which we read how the photon temperature is modified in the presence of acoustic reheating:
\begin{equation}
T_{\gamma}(a)=T_{\gamma,0}\exp \left [-\delta\vartheta(a) \right ] / a\,.
\label{eq:tgamma}
\end{equation}
As $T_{\gamma,0}$ is fixed to high precision by COBE-FIRAS observations and serves as boundary condition for $T_{\gamma}(a)$, whereas $\delta\vartheta(a>a_0)$ is strictly positive, the photon bath is colder in the past relative to the standard $T_{\gamma} \propto a^{-1}$ evolution: this reflects how acoustic reheating causes photons to cool less rapidly compared to the standard thermal history (see Fig.~\ref{fig:temperature_evolution} for a schematic illustration).

Decoupling of neutrinos occurs at $T \sim 2\,{\text{MeV}}$, when the scale factor was $a_{\text{dec}}$~\cite{EscuderoAbenza:2020cmq,Bennett:2020zkv,Escudero:2025kej}. For $a>a_{\text{dec}}$, since neutrinos do not take part in the entropy injection process, their temperature evolves as $T_{\nu}\propto a^{-1}$. Fixing the value of the CMB temperature today, a straightforward comoving entropy conservation argument implies the following value for the neutrino temperature today:
\begin{equation}
T_{\nu,0} = \left ( \frac{4}{11} \right ) ^{\frac{1}{3}} T_{\gamma,0} \exp \left [ -\delta\vartheta(a_{\text{dec}}) \right ] \,,
\label{eq:tnu}
\end{equation}
suppressed by a factor $\exp[-\delta\vartheta(a_{\rm dec})]$ relative to the standard thermal history value. Equivalently, the relic neutrino number density is suppressed by a factor $\exp[-3\delta\vartheta(a_{\text{dec}})]$, since $n_{\nu,0}\propto T_{\nu,0}^3$, directly impacting CNB capture rates. Only diffusion damping occurring after neutrino decoupling contributes to the effect, as earlier entropy injection is shared by photons and neutrinos. This restricts our small-scale reach to $k \lesssim k_D(a_{\rm dec}) \approx 3 \times 10^5\,{\text{Mpc}}^{-1}$, corresponding to modes whose damping occurs after neutrino decoupling, whereas modes at larger wavenumbers dissipate while neutrinos are coupled, and therefore do not affect $T_{\nu}/T_{\gamma}$~\cite{Jeong:2014gna,Nakama:2014vla}.

\textbf{\textit{Models for small-scale power.}} --- A wide class of early Universe models predicts significant departures from near scale-invariance beyond the narrow CMB--LSS window. Since acoustic reheating depends on the integrated small-scale power entering $\langle\Theta^2(a)\rangle$, it is neither necessary nor meaningful to focus on specific models and associated spectra. To stay as model-agnostic as possible, we consider three templates describing different types of theoretically motivated small-scale features.

We first study \textit{localized enhancements} of $\Delta_{\cal R}^2(k)$, considering a sharp \textit{monochromatic peak}, which is an idealized description of models where perturbations are amplified on a very narrow range of scales. This can arise in models featuring brief violations of slow-roll or non-attractor phases, turns in multi-field inflationary trajectories, and particle production~\cite{Barnaby:2009dd,Garcia-Bellido:2017mdw,Ballesteros:2022hjk,Ozsoy:2023ryl,Wang:2025hwc}, and is a useful and widely adopted benchmark~\cite{Ananda:2006af,NANOGrav:2023hvm}. We parametrize this as $\Delta_{\cal R}^2(k) = \alpha k_{\star}\delta(k-k_{\star})$, with $\alpha$ controlling the amplitude of the enhancement at the scale $k_{\star}$.

Amplification phases of finite duration are naturally captured by a \textit{broad peak} template. Such phases occur e.g.\ in single-field models with inflection points leading to ultra-slow-roll dynamics during which curvature perturbations are amplified. As these dynamics usually persist over a finite number of $e$-folds, the enhancement is localized in $\ln k$, and can be parametrized via the widely used lognormal template~\cite{Germani:2017bcs,Motohashi:2017kbs,Ballesteros:2017fsr,Ianniccari:2024bkh,Pi:2024ert}, routinely used in the context of PBH formation and scalar-induced GWs (SIGWs)~\cite{NANOGrav:2023hvm,Cecchini:2025oks}. We denote the amplitude, position, and width (in $\ln k$) of the template by $\alpha$, $k_{\star}$, and $\sigma$.

Complementary to localized features are \textit{extended enhancements} of the small-scale spectrum. The simplest choice is a \textit{blue-tilted power-law} on scales beyond the CMB--LSS ones, $\Delta_{\cal R}^2(k) = \alpha(k/k_{\star})^{n_{\star}}$, where $n_{\star}>0$ is the power-law index, and we set the pivot scale $k_{\star}=k_D(a_{\rm dec})$, so $\alpha$ is the amplitude at the neutrino decoupling diffusion scale. These enhancements naturally occur in various scenarios, such as non-attractor inflation, models with excited initial states, or in the presence of spectator fields whose fluctuations later convert into curvature perturbations~\cite{Kawasaki:2013xsa,Ballesteros:2018wlw,Franciolini:2018vbk}. The $n_{\star}=3$ case gives a white noise contribution to $P(k)$, generated for instance by Poisson fluctuations from discrete sources.

\textbf{\textit{PTOLEMY forecasts.}} --- PTOLEMY aims to detect the CNB via neutrino capture on tritium ($^3H$), with the energy spectrum of the emitted electron ($d\Gamma/dE_e$) being the primary observable. The spectrum is dominated by $^3H$ $\beta$-decay up to the $\beta$-decay endpoint, followed by a narrow peak associated to relic neutrino capture, and smeared by finite experimental energy resolution~\cite{PTOLEMY:2019hkd}. Under optimistic assumptions, the expected number of capture events is at most ${\cal O}(10)$ per year, so the signal lies within a few energy bins near the endpoint~\cite{Cocco:2007za,PTOLEMY:2019hkd}.

For our purposes, as only the normalization of the event rate is affected, detailed spectral information carries little statistical weight, so the relevant observable reduces to the observed number of capture events in a narrow region of the spectrum just above the $\beta$-decay endpoint. We therefore focus on an energy window $[E_1,E_2]=[E_{\text{cap}}-\Delta_E,E_{\text{cap}}+\Delta_E]$ centered on the capture feature $E_{\text{cap}}$, with $\Delta_E$ set by the experimental energy resolution. Following the PTOLEMY collaboration sensitivity analysis, we take $\Delta_E=10\,{\text{meV}}$ as an optimistic target resolution~\cite{PTOLEMY:2019hkd}. We further assume that backgrounds in this window can be controlled to a level where the measurement is statistics-limited. While this assumption is experimentally challenging and non-trivial, we adopt it as an idealized baseline to study how acoustic reheating effects can manifest in PTOLEMY, rather than provide a fully-fledged experimental forecast. We conservatively set the local CNB clustering factor $f_c$ to unity, i.e.\ we ignore the gravitational enhancement of the CNB. Since clustering would only increase the capture rate and hence our sensitivity, and there is currently no consensus in the literature on the size of $f_c$, setting $f_c=1$ yields conservative estimates without committing to specific astrophysical models~\cite{Singh:2002de,Ringwald:2004np,Zhang:2017ljh,Mertsch:2019qjv,Bondarenko:2023ukx}. We effectively (conservatively) assume neutrino masses close to the minimal values, consistently with current very tight cosmological upper limits~\cite{Vagnozzi:2017ovm,Vagnozzi:2018jhn,Giusarma:2018jei,RoyChoudhury:2019hls,Tanseri:2022zfe,Green:2024xbb,Naredo-Tuero:2024sgf,Du:2024pai,Jiang:2024viw,RoyChoudhury:2024wri,Escudero:2024uea,RoyChoudhury:2025dhe,Zhou:2025nkb,RoyChoudhury:2025iis,Feng:2026pzs}. Finally, we assume a large effective $^3H$ target mass $M_T$, and long exposure time $T_{\text{exp}}$, so the expected number of events is limited primarily by statistics rather than instrumental reach.

Under these idealized, simplifying assumptions, we can compute the total number of CNB events in the signal window $\overline{N}$ as a function of the model parameters, collectively denoted by $\boldsymbol{\theta}$. We first define the benchmark number of events $\overline{N}_0$ expected in the standard thermal history case, in the absence of acoustic reheating~\cite{PTOLEMY:2019hkd}:
\begin{equation}
\overline{N}_0 \equiv \frac{T_{\rm exp}}{\sqrt{2\pi}\sigma} \int_{E_1}^{E_2}dE'\,\frac{d\Gamma}{dE_e}(E')\exp \left [ -\frac{(E_e-E')^2}{2\sigma^2} \right ] \,,
\label{eq:benchmarkn}
\end{equation}
where $\sigma \equiv \Delta_E/\sqrt{8\log 2}$, $d\Gamma/dE_e$ is the differential CNB capture rate, whose expression is well-known in the literature and scales linearly with $M_T$~\cite{PTOLEMY:2019hkd}. We assume $M_T=100\,{\text{g}}$ and $T_{\text{exp}}=10\,{\text{yr}}$, both optimistic but potentially achievable benchmarks. For a theoretical small-scale power spectrum model specified by parameters $\boldsymbol{\theta}$, the relic neutrino number density is modified while leaving the capture rate unchanged, leading to a uniform rescaling of the expected number of events:
\begin{equation}
\overline{N}(\boldsymbol{\theta})=\overline{N}_0 \exp \left [ -3 \delta\vartheta(\boldsymbol{\theta}) \right ] \,.
\end{equation}
where the dependence on $\boldsymbol{\theta}$ enters entirely through $\delta\vartheta$, and therefore through $\langle\Theta^2\rangle$ as given in Eq.~(\ref{eq:theta}). For our purposes the measurement is effectively a counting experiment, so the likelihood of observing $N$ events is modeled using a Poisson likelihood with mean given by the theoretical expectation $\overline{N}(\boldsymbol{\theta})$~\cite{Alvey:2021xmq}. We then assume that the observed event count coincides with the standard thermal history expectation, $N=\overline{N}_0$, corresponding to a ``null detection'' of acoustic reheating.

\textbf{\textit{Results.}} --- We forecast limits on the parameters of the small-scale models introduced earlier, under the previously discussed ``null detection'' assumption. In other words, we ask: \textit{if PTOLEMY observes a capture rate consistent with the standard expectation of $336\,{\text{cm}}^{-3}$, what constraints can be placed on small-scale primordial fluctuations beyond the reach of the CMB and LSS?} All bounds are quoted at $68\%$ confidence level.

For the monochromatic peak, provided the peak lies below the diffusion scale at neutrino decoupling $k_D(a_{\rm dec}) \approx 3 \times 10^5\,{\text{Mpc}}^{-1}$, acoustic reheating only depends on the total energy injected after decoupling, rather than the peak's location. We find that a null detection by PTOLEMY would set a limit $\alpha \lesssim 0.06$.

For the lognormal template, we forecast comparable constraints on the overall amplitude, $\alpha \lesssim 0.09$. The fact that this bound is not significantly weakened despite the additional width parameter $\sigma$ reflects the fact that it is the integrated small-scale power rather than detailed spectral shape that matters. A null detection by PTOLEMY would also set a limit of $\sigma \lesssim 2.5$: larger values would spread the feature over a larger range of modes, leading to stronger effects. From the usual correspondence $\ln k \simeq N$ at horizon exit during inflation, this would exclude mechanisms responsible for the feature persisting for more than $N \lesssim {\cal{O}}(3)$ $e$-folds.

For the blue power-law template, we find $\alpha \lesssim 0.09$ and $n_{\star} \lesssim 5$: larger values of $n_{\star}$ lead to a more rapid increase of power, implying stronger effects and a larger suppression of the relic neutrino abundance. The precise interpretation of this bound would depend on the specific mechanism responsible for the enhancement.

Summing up, our results show that, even across different templates, a null detection by PTOLEMY would constrain the amplitude of the small-scale power spectrum $\alpha$ to the ${\cal O}(0.1)$ for $k \lesssim 3 \times 10^5\,{\text{Mpc}^{-1}}$, i.e.\ those whose dissipation occurs after neutrino decoupling. This level of sensitivity can be understood from counting statistics. For ${\cal O}(100)$ events (over 10 years), Poisson statistics imply a relative uncertainty of $1/\sqrt{N} \sim 0.1$, while the expected rate scales as $\overline{N} \simeq \overline{N}_0(1-3\delta\vartheta)$ for small $\delta\vartheta$, implying that $\delta\vartheta$ is constrained to a comparable level. Since $\delta\vartheta$ is proportional to $\langle\Theta^2\rangle$ and thus to $\alpha$ for all templates, ${\cal O}(0.1)$ limits on $\alpha$ are expected.

\textbf{\textit{Discussion.}} --- We have shown that the science reach of PTOLEMY (e.g.\ Refs.~\cite{Huang:2016qmh,Horvat:2017gfm,Horvat:2017aqf,Bondarenko:2020vta,Alvey:2021sji,Das:2022xsz,Capolupo:2022hhr,Banerjee:2023lrk,Martinez-Mirave:2024dmw,Lambiase:2025twn}) extends well beyond CNB detection, as the experiment is indirectly sensitive to the primordial power spectrum for $k \lesssim 3 \times 10^5\,{\text{Mpc}}^{-1}$, through the dissipation of the relevant modes after neutrino decoupling~\cite{Chluba:2012gq}. The effect depends only on the integral of the small-scale spectrum across all modes dissipating after neutrino decoupling, rather than on its detailed shape. This is why, across the three templates considered, our limits are relatively insensitive to the precise spectral shape. Our results can be therefore be interpreted as limits on an effective amplitude of small-scale power. Since the relevant modes are erased well before recombination, acoustic reheating is a particularly clean probe of small-scale power, determined only by early-Universe dissipation in the perturbative regime.

\begin{figure*}
\centering
\includegraphics[width=0.44\textwidth]{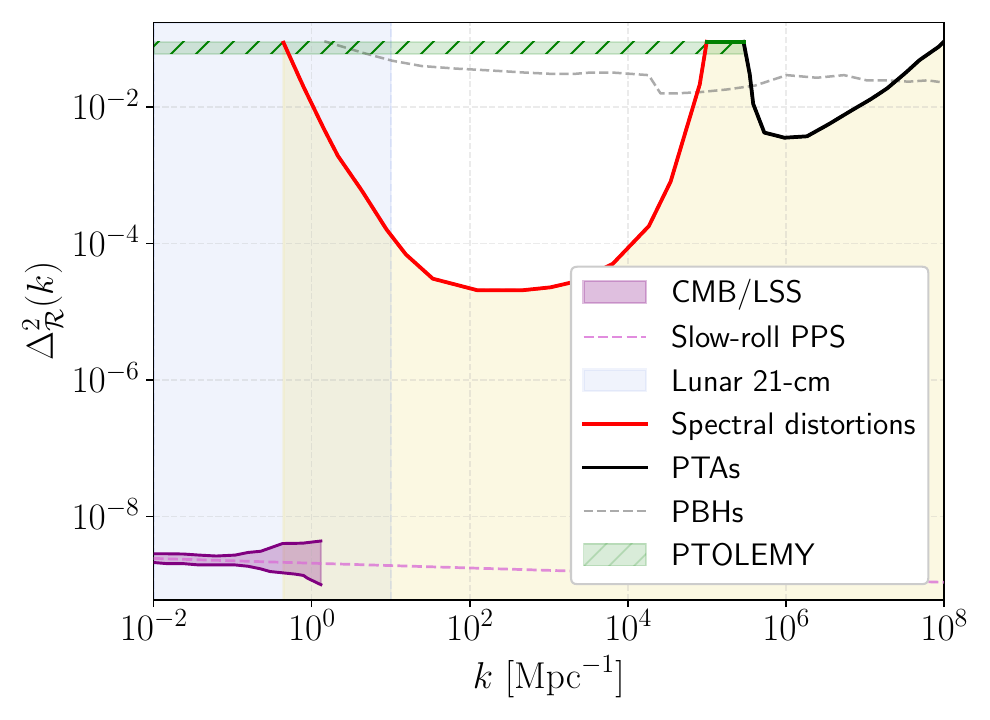}\hfill
\includegraphics[width=0.44\textwidth]{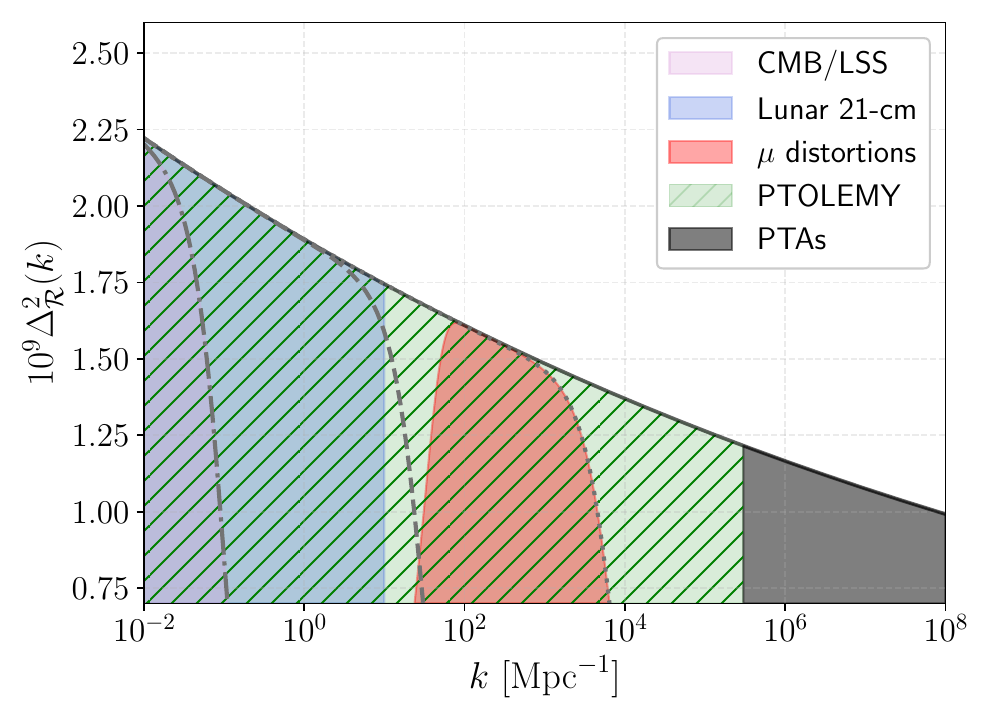}
\caption{\textit{Left}: constraints on the primordial power spectrum from CMB and LSS (purple), spectral distortions (red), PTAs (black), and PBHs (gray). The hatched green band shows projected PTOLEMY upper limits, and the blue band is the range accessible to future Lunar 21-cm observations~\cite{Cole:2019zhu}. The khaki shaded region is the parameter space allowed beyond the CMB--LSS region by existing constraints, and the dashed magenta curve shows the near scale-invariant spectrum expected from slow-roll inflation. \textit{Right}: schematic illustration, inspired by Fig.~1 of Ref.~\cite{Pajer:2012vz}, of the scales probed by the observables on the left, with the red region showing the cumulative contribution of Silk-damped modes to $\mu$-type distortions.}
\label{fig:constraints_pps}
\end{figure*}

It is useful to compare our forecasts to existing and future probes of small-scale power, see Fig.~\ref{fig:constraints_pps}. Light element abundances constrain entropy injection after BBN. Since BBN takes place shortly after neutrino decoupling, these bounds probe a similar range of modes, up to $k \lesssim 10^5\,{\text{Mpc}}^{-1}$, allowing enhancements in $\Delta_{\cal R}^2(k)$ at the $10^{-2}$--$10^{-1}$ level~\cite{Jeong:2014gna,Nakama:2014vla}. However, such limits rely on inferred primordial light element abundances, and are affected by significant astrophysical uncertainties, especially in the determination of the primordial Helium fraction. On the other hand, CMB SDs are sensitive to the dissipation of $10 \lesssim k/{\text{Mpc}}^{-1} \lesssim 10^4$ modes. In this range, SDs constrain $\Delta_{\cal R}^2(k)$ to the $10^{-4}$--$10^{-5}$ level~\cite{Chluba:2012we}.

The above limits have been derived assuming a top-hat template: we have not adopted this here, as it does not naturally arise in well-motivated early Universe scenarios. Still, order-of-magnitude comparisons between PTOLEMY, BBN, and SDs are meaningful, as acoustic reheating is sensitive to integrated dissipation of small-scale power, rather than detailed spectral features. In this context, relic neutrinos are an independent and clean probe which, while weaker than BBN, avoids astrophysical systematics inherent to primordial abundance measurements, thereby offering a valuable consistency test.

Pulsar timing arrays (PTAs) probe enhancements on smaller scales via limits on SIGWs~\cite{Domenech:2021ztg}. For instance, NANOGrav can probe modes $10^6 \lesssim k/{\text{Mpc}}^{-1} \lesssim 10^8$, just beyond PTOLEMY's range, and the NANOGrav signal can be explained by enhanced small-scale power modeled either via the monochromatic or lognormal templates, with amplitudes $\Delta_{\cal R}^2 \gtrsim 0.03$~\cite{NANOGrav:2023hvm}. PTAs and relic neutrinos are therefore complementary probes, relying on different mechanisms and probing different scales.

Finally, strong limits are also often quoted from the non-observation of PBHs and ultracompact minihalos (UCMHs)~\cite{Scott:2009tu,Bringmann:2011ut,Yang:2013dsa,Cole:2017gle,Nakama:2017qac,Sato-Polito:2019hws,FrancoAbellan:2023sby,Gouttenoire:2025wxc,Bringmann:2025cht,Carr:2026hot}. These can exclude amplitudes $\Delta_{\cal R}^2 \gtrsim 10^{-5}$ up to $k \sim 10^8\,{\text{Mpc}}^{-1}$ (and potentially even smaller scales). These limits are subject to significant theoretical uncertainties, as they depend on the collapse prescription relating primordial perturbations to collapsed objects, and assumptions on the treatment of non-Gaussianity and quantum diffusion~\cite{Gow:2020bzo,DeLuca:2025nao}, which can alter the limits by a few orders of magnitude. While weaker, our limits provide a different probe of small-scale power, relying only on well-understood early Universe physics, rather than the formation of rare non-linear objects.

Beyond the adiabatic perturbations studied here, our considerations can be applied to isocurvature modes. For matter isocurvature perturbations, one can still obtain an expression analogous to Eq.~(\ref{eq:theta}) relating $\langle\Theta^2(a)\rangle$ to the dimensionless isocurvature power spectrum $\Delta_{\cal S}^2$. As a possible application, PBHs contribute a Poissonian matter isocurvature component with $\Delta_{\cal S}^2(k) \propto k^3f_{\text{pbh}}^2/n_{\text{pbh}}$ (i.e.\ the familiar Poisson plateau in the dimensional isocurvature power spectrum), where $f_{\text{pbh}}$ is the fraction of dark matter in the form of PBHs, and $n_{\text{pbh}}$ is the PBH comoving number density~\cite{Meszaros:1975ef,Afshordi:2003zb,Gong:2017sie,Carr:2018rid,Murgia:2019duy}. Assuming a monochromatic PBH mass function with mass $M_{\text{pbh}}$, so $n_{\text{pbh}} \propto f_{\text{pbh}}\rho_{\text{dm}}/M_{\text{pbh}}$, the acoustic reheating signal is expected to depend on the combination $f_{\text{pbh}}M_{\text{pbh}}$. These considerations are intended at the level of scaling arguments, with a detailed study left to follow-up work.

Aside from acoustic reheating, relic neutrinos provide information complementary to CMB and BBN limits on the relativistic energy density $\rho_r$. These constrain the total radiation energy density via $N_{\text{eff}}$, whereas the CNB is sensitive to $n_{\nu,0}$. Assuming thermal neutrino spectra and zero chemical potentials, $\rho_r$ and $n_{\nu,0}$ are related via $T_{\nu,0}$. Comparing $n_{\nu,0}$ and CMB/BBN limits on $N_{\text{eff}}$  then constitutes a non-trivial consistency test of the neutrino thermal history. This could also test scenarios such as late entropy injection or low-scale reheating, which can lower the neutrino contribution to $N_{\text{eff}}$~\cite{Kawasaki:1999na,Hannestad:2004px,deSalas:2015glj,Gerbino:2016sgw,Vagnozzi:2019ezj,Vagnozzi:2020gtf,Giovanetti:2021izc,Vagnozzi:2023lwo,Oikonomou:2023qfz,Ben-Dayan:2023lwd,Odintsov:2024grb,Barbieri:2025moq,Ben-Dayan:2025bqd,Escudero:2025avx,Escudero:2026mgw}.

In summary, relic neutrinos can set qualitatively new limits on primordial perturbations beyond the CMB--LSS scales. These limits are not specific to inflation, but apply to any mechanism generating primordial perturbations. The modes of interest were previously considered inaccessible to direct probes, as diffusion damping erases them well before recombination. Relic neutrinos thus emerge not only as a near-future detection target~\cite{PTOLEMY:2019hkd}, but also as a complementary probe of primordial physics.

\textbf{\textit{Acknowledgments.}} --- S.V.\ acknowledges helpful discussions with Will Coulton and Atsuhisa Ota. S.V.\ acknowledges support from INFN through Iniziativa Specifica ``Quantum Fields in Gravity, Cosmology and Black Holes'' (FLAG). This publication is based upon work from the COST Action CA21136 ``Addressing observational tensions in cosmology with systematics and fundamental physics'' (CosmoVerse), supported by COST.

\bibliography{PTOLEMYfundamentalphysics}

\end{document}